\documentclass[12pt]{article}

\usepackage{scicite}

\usepackage{times}

\usepackage[pdftex]{graphicx}
\usepackage{pdfpages}

\newenvironment{sciabstract}{%
\begin{quote} \bf}
{\end{quote}}

\newcounter{lastnote}
\newenvironment{scilastnote}{%
\setcounter{lastnote}{\value{enumiv}}%
\addtocounter{lastnote}{+1}%
\begin{list}%
{\arabic{lastnote}.}
{\setlength{\leftmargin}{.22in}}
{\setlength{\labelsep}{.5em}}}
{\end{list}}

\title{Tunable Fractional Quantum Hall Phases in Bilayer Graphene}

\author
{Patrick Maher,$^{1\dag}$ Lei Wang,$^{2\dag}$ Yuanda Gao,$^{3}$ Carlos Forsythe,$^{1}$ Takashi Taniguchi,$^{4}$\\
Kenji Watanabe,$^{4}$ Dmitry Abanin,$^{5,6}$ Zlatko Papi\'c,$^{5,6}$ Paul Cadden-Zimansky,$^{7}$ \\
James Hone,$^{3}$ Philip Kim$^{1\ast}$ and Cory R. Dean$^{8\ast}$
\\
\normalsize{$^{1}$Department of Physics, Columbia University, New York, NY 10027, USA}\\
\normalsize{$^{2}$Department of Electrical Engineering, Columbia University, New York, NY 10027, USA}\\
\normalsize{$^{3}$Department of Mechanical Engineering, Columbia University, New York, NY 10027, USA}\\
\normalsize{$^{4}$National Institute for Materials Science, 1-1 Namiki, Tsukuba, Japan}\\
\normalsize{$^{5}$Perimeter Institute for Theoretical Physics, Waterloo, ON N2L 2Y5, Canada}\\
\normalsize{$^{6}$Institute for Quantum Computing, Waterloo, ON N2L 3G1, Canada}\\
\normalsize{$^{7}$Physics Program, Bard College, Annandale-on-Hudson, NY 12504, USA}\\
\normalsize{$^{8}$Department of Physics, The City College of New York, New York, NY 10031, USA}\\
\\
\normalsize{$^\dag$These authors contributed equally to this work.}
\\
\normalsize{$^\ast$Corresponding author. E-mail:  pk2015@columbia.edu (P.K.) and cdean@ccny.cuny.edu (C.R.D.)}
}

\date{}

\begin{document} 


\baselineskip24pt


\maketitle


\begin{sciabstract}
Symmetry breaking in a quantum system often leads to complex emergent behavior. In bilayer graphene (BLG), an electric field applied perpendicular to the basal plane breaks the inversion symmetry of the lattice, opening a band gap at the charge neutrality point. In a quantizing magnetic field electron interactions can cause spontaneous symmetry breaking within the spin and valley degrees of freedom, resulting in quantum Hall states (QHS) with complex order. Here we report fractional quantum Hall states (FQHS) in bilayer graphene which show phase transitions that can be tuned by a transverse electric field. This result provides a model platform to study the role of symmetry breaking in emergent states with distinct topological order.
\end{sciabstract}

The fractional quantum Hall effect\cite{Tsui_two-dimensional_1982} (FQHE) represents a spectacular example of emergent behavior where strong Coulomb interactions drive the existence of a correlated many body state. In conventional III-V heterostructures, the celebrated Laughlin wave function\cite{Laughlin_anomalous_1983} together with the composite fermion picture\cite{Jain_composite-fermion_1989} provides a complete description of the series of  FQHE states that have been observed within the lowest Landau level.  In higher Landau levels the situation remains less clear, as in addition to Laughlin states, new many-body phases appear such as the still  controversial even-denominator 5/2 state\cite{Willett_observation_1987} (presumed to be a Pfaffian with non-abelian quantum statistics\cite{moore_nonabelions_1991}), and a variety of charge-density wave states\cite{Koulakov_charge_1996,Moessner_exact_1996}.

Recently the nature of the FQHE in graphene has received intense interest\cite{Bolotin_observation_2009,Du_fractional_2009,Bao_magnetoconductance_2010,Dean_multicomponent_2011,Feldman_unconventional_2012,Ki_even_2013,Kou_electron-hole_2013} since the combined spin and valley degrees of freedom are conjectured to yield novel FQHE states within an approximate SU(4) symmetry space (assuming relatively weak spin Zeeman and short-range interaction energies can be ignored). Furthermore, unlike conventional semiconductor systems, which have shown limited evidence for transitions in fractional states\cite{Eisenstein_Evidence_1989,Du_fractional_1995,kang_evidence_1997,lai_fractional_2004,liu_stability_2011}, the wide gate tunability of graphene systems coupled with large cyclotron energies allows for the exploration of multiple different SU(4) order parameters for a large range of filling fractions. In bilayer graphene, the capability to force transitions between different spin and valley orderings by coupling to electric fields perpendicular to the basal plane as well as magnetic fields provides a unique opportunity to probe interaction-driven symmetry breaking within this expanded SU(4) basis in a fully controllable way\cite{Weitz_broken-symmetry_2010,Kim_spin-polarized_2011,Velasco_jr_transport_2012,Maher_Evidence_2013}. The most intriguing consequence of this field tunability is the possibility to induce transitions between different FQHE phases\cite{Apalkov_controllable_2010,Papic_tunable_2011,Snizhko_importance_2012}. Thus, BLG provides a unique model system to experimentally study phase transitions between different topologically-ordered states.

While observation of the FQHE in monolayer graphene has now been reported in several studies\cite{Bolotin_observation_2009,Du_fractional_2009,Dean_multicomponent_2011,Feldman_unconventional_2012} including evidence of magnetic-field-induced phase transitions\cite{Feldman_fractional_2013}, achieving the necessary sample quality in bilayer graphene (BLG) has proven a formidable challenge\cite{Bao_magnetoconductance_2010,Ki_even_2013,Kou_electron-hole_2013}. Here we fabricate BLG devices encapsulated in hexagonal boron nitride using the recently developed van der Waals transfer technique\cite{Wang_one-dimensional_2013} (see supplementary materials (SM) section 1.1). The device geometry includes both a local graphite bottom gate and an aligned metal top gate, which allows us to independently control the carrier density in the channel ($n=(C_{TG}V_{TG}+C_{BG}V_{BG})/e-n_0$, where $C_{TG}$ ($C_{BG}$) is the top (bottom) gate capacitance per area, $V_{TG}$ ($V_{BG}$) is the top (bottom) gate voltage, $e$ is the electron charge, and $n_0$ is residual doping) and the applied average electric displacement field ($D=(C_{TG}V_{TG}-C_{BG}V_{BG})/2\epsilon_0-D_0$, where $D_0$ is a residual displacement field due to doping). Crucially, these devices have the portion of the graphene leads that extend outside of the dual-gated channel exposed to the silicon substrate, which we utilize as a third gate to set the carrier density of the leads independently (Fig. 1a). We have found that tuning the carrier density in the graphene leads has a dramatic effect on the quality of magnetotransport data (see SM section 2.2) allowing us to greatly improve the quantum Hall signatures, especially at large applied magnetic fields. Due to a slight systematic n-doping of our contacts during fabrication, our highest-quality data is obtained  for an n-doped channel. We thus restrict our study to the electron side of the band structure.

Under application of low magnetic fields, transport measurements show a sequence of QHE plateaus in $R_{xy}$ appearing at $h/4me^2 $, where $m$ is a non-zero integer, together with resistance minima in $R_{xx}$, consistent with the single-particle Landau level spectrum expected for bilayer graphene\cite{McCann_landau-level_2006}. At fields larger than $\sim$5~T, we observe complete symmetry breaking with QHE states appearing at all integer filling fractions, indicative of the high quality of our sample (fig. 1b). By cooling the sample to sub-kelvin temperatures (20-300 mK) and applying higher magnetic fields (up to 31~T), clearly developed fractional quantum Hall states (FQHSs) appear at partial LL filling, with vanishing $R_{xx}$ and unambiguous plateaus in $R_{xy}$. Fig. 1d shows an example of a remarkably well formed FQHS at LL filling fraction $\nu=2/3$ appearing at approximately 25~T. By changing  $V_{TG}$ while sweeping $V_{BG}$, we can observe the effect of  different displacement fields on these FQHSs. Fig.~1d shows an example of this behaviour at two demonstrative top gate voltages. For $V_{TG}=0.2$~V, the $\nu=2/3$ and $\nu=5/3$ FQHSs are clearly visible as minima $\sigma_{xx}$, whereas for $V_{TG}=1.2$~V, the $\nu=2/3$ state is completely absent, the 5/3 state appears weakened, and a new state at $\nu=4/3$ becomes visible. This indicates that both the existence of the FQHE in BLG, and importantly the sequence of the observed states, depend critically on the applied electric displacement field, and that a complete study of the fractional hierarchy in this material requires the ability to independently vary the carrier density and displacement field. 

To more clearly characterize the effect of displacement field it is illuminating to remap the conductivity data versus displacement field and LL filling fraction $\nu$. One such map is shown in fig. 2a where we have focused on filling fractions between $0<\nu<4$. Replotted in this way a distinct sequence of transitions, marked by compressible regions with increased conductivity, is observed for each LL.  For example, at $\nu=1$ there is evidently a phase transition exactly at $D=0$, and then a second transition at large finite $D$. By contrast at $\nu=2$, there is no apparent transition at $D=0$, and while there is a transition at finite $D$, it appears at much smaller displacement field than at $\nu=1$. Finally at $\nu=3$ there is a single transition only observed at $D=0$. This pattern is in agreement with other recent experiments\cite{lee_chemical_2014}. At large displacement fields it is expected that it is energetically favorable to maximize layer polarization, indicating that low-displacement-field states which undergo a transition into another state at large displacement field (e.g. $\nu=1,2$) likely exhibit an ordering different from full layer polarization. Following predictions that polarization in the 0-1 orbital degeneracy space is energetically unfavorable\cite{Barlas_intra-landau-level_2008}, this could be a spin ordering, like ferromagnetism or antiferromagnetism, or a layer-coherent phase\cite{Barlas_intra-landau-level_2008,gorbar_broken-symmetry_2011,roy_theory_2012,lambert_quantum_2013}. This interpretation is consistent with several previous experimental studies which reported transitions within the symmetry broken integer QHE states to a layer polarized phase under finite displacement field\cite{Weitz_broken-symmetry_2010,Kim_spin-polarized_2011,Velasco_jr_transport_2012,Maher_Evidence_2013,lee_chemical_2014}.  

At higher magnetic field and lower temperature (Fig.~2b) the integer states remain robust at all displacement fields, indicating that the observed transitions in Fig.~2a are actually continuous non-monotonic gaps that are being washed out by temperature or disorder when the gap is small. At these fields FQH states within each LL become evident, exhibiting transitions of their own with displacement field. Of particular interest in our study are the states at $\nu=2/3$ and $\nu=5/3$ which are the most well developed. A strong $\nu=2/3$ state is consistent with recent theory\cite{Papic_topological_2014}, which predicts this state to be fully polarized in orbital index in the 0 direction. At $\nu=2/3$, there is a clear transition apparent at $D=0$, as well as two more at $|D|\approx 100$~mV/nm. These transitions are qualitatively and quantitatively similar to the transitions in the $\nu=1$ state seen in Fig.~2b. For the 5/3 state, we present high-resolution scans at fields from 20-30~T in Fig.~2c. Here we again observe a transition at $D=0$ as well as at finite $D$. The finite $D$ transitions, however, occur at much smaller values than for either $\nu=2/3$ or 1. Indeed, they are much closer in $D$ value to the transitions taking place at $\nu=2$, which is almost a factor of $\sim$8 smaller than that of $\nu=1$. 

In Fig.~3a-b we plot the resistance minima of 2/3 and 5/3 state, respectively, as a function of $D$ (corresponding to vertical line cuts through the 2D map in Fig. 2b). For the 2/3 state we observe a broad transition, whose position we define by estimating the middle of the transition. The 5/3 state exhibits much narrower transitions, which we mark by the local maximum of resistance. Fig.~3c plots the location of transitions in $D$ as a function of magnetic field $B$ for both integer and fractional filling fractions. Our main observation is that the transitions in the fractional quantum Hall states and the transitions in the parent integer state (i.e., the smallest integer larger than the fraction) fall along the same line in $D$ vs. $B$. More specifically, the finite $D$ transitions for $\nu=2/3$ and $\nu=1$ fall along a single line of slope 7 mV/nm$\cdot$T, and the transitions for $\nu=5/3$ and $\nu=2$ fall along a single line of slope 0.9 mV/nm$\cdot$T. 

We now turn to a discussion of the nature of these transitions. For the lowest LL of BLG, the possible internal quantum state of electrons comprises an octet described by the SU(4) spin-valley space and the 0-1 LL orbital degeneracy. Since $\nu=1$ corresponds to filling 5 of the 8 degenerate states in the lowest LL, the system will necessarily be polarized in some direction of the SU(4) spin-valley space. The presence of a $D=0$ transition for $\nu=1$ indicates that even at low displacement field, the ground state exhibits a layer polarization which changes as $D$ goes through zero. At the same time, we expect that at large $D$ the system is maximally layer polarized.  We therefore propose that the transition at finite $D$ is between a 1/5 layer-polarized state (e.g. 3 top layer, 2 bottom layer levels filled) and a 3/5 layer-polarized state (e.g. 4 top layer, 1 bottom layer levels filled)\cite{gorbar_broken-symmetry_2011,roy_theory_2012,lambert_quantum_2013}. Interestingly, the quantitative agreement of the transitions for $\nu=2/3$ and $\nu=1$, shown in Fig.~3c, strongly suggest that the composite fermions undergo a similar transition in layer polarization.

While the observed transition at $\nu=1$ and its associated FQHE (i.e., $\nu=2/3$) can be explained by a partial-to-full layer polarization transition, the nature of the transition at $\nu=2$ is less clear: presumably again the high $D$ state exhibits layer polarization, but we do not have any experimental insight as to the ordering of the low $D$ state. Additionally, while the finite $D$ transitions in the 5/3 state seem to follow the $\nu=$2 transitions quantitatively (Fig.~3c), the 5/3 state also has a clear transition at $D=0$, suggesting there may be a different ground state ordering of the 5/3 and 2 states very near $D=0$. In particular, the transition at $D=0$ indicates that the $\nu=5/3$ state exhibits layer polarization even at low displacement field, in a state separate from the high-displacement-field layer polarized state, whereas this does not appear to be true at $\nu=2$.  One possible explanation for this result could be the formation of a layer-coherent phase that forms at $\nu=2$ filling\cite{lambert_quantum_2013}, but which is not stabilized at partial filling of the Landau level. 

We also briefly mention other observed FQHSs. At the highest fields we see evidence of $\nu=1/3$ and $\nu=4/3$ (see supplemental 2.3 and Fig. 2c), with both exhibiting a phase transitions that appear to follow those of $\nu=2/3$ and $\nu=5/3$, respectively. We have also observed transitions in the $\nu=8/3$ state with displacement field (see Fig. 2b), though we do not have a systematic study of its dependence on magnetic field due to our limited gate range. Qualitatively, there is a region at low $D$ where no minimum is apparent which gives way to a plateau and minimum at finite $D$. The $\nu=3$ state exhibits a similar transition at $D=0$, indicating that there may also be a correspondence between the integer and fractional states in the $\nu=$3 Landau level. The inset of Fig.~3c summarizes these observations, while the detailed data are available in SM section 2.6. Lastly, we have preliminary evidence that the fractional hierarchy breaks electron-hole symmetry\cite{Kou_electron-hole_2013} (see SM section 2.3), as the clearest fractional states we observe can be described as $\nu=m-1/3$ where $m$ is an integer. 

The electric-field-driven phase transitions observed in BLG's FQHE indicate that ordering in the SU(4) degeneracy space is critical to the stability of the FQHE. In particular, quantitative agreement between transitions in FQH states and those in parent integer QH states suggests that generally the composite fermions in BLG inherit the SU(4) polarization of the integer state, and couple to symmetry breaking terms with the same strength. However, an apparent disagreement in the transition structure at $\nu=$5/3 and $\nu=$2 indicates that there may be subtle differences in the ground state ordering for the integer and fractional quantum Hall states.

\bibliographystyle{Science}

\begin{scilastnote}
\item  We acknowledge Minkyung Shinn and Gavin Myers for assistance with measurements. We thank Amir Yacoby for helpful discussions. A portion of this work was performed at the National High Magnetic Field Laboratory, which is supported by US National Science Foundation cooperative agreement no. DMR-0654118, the State of Florida and the US Department of Energy. This work is supported by the National Science Foundation (DMR-1124894) and FAME under STARnet. P.K acknowledges support from DOE (DE-FG02-05ER46215).
\end{scilastnote}

\clearpage

\begin{figure}[t]
    \begin{center}
	\includegraphics[width=\linewidth]{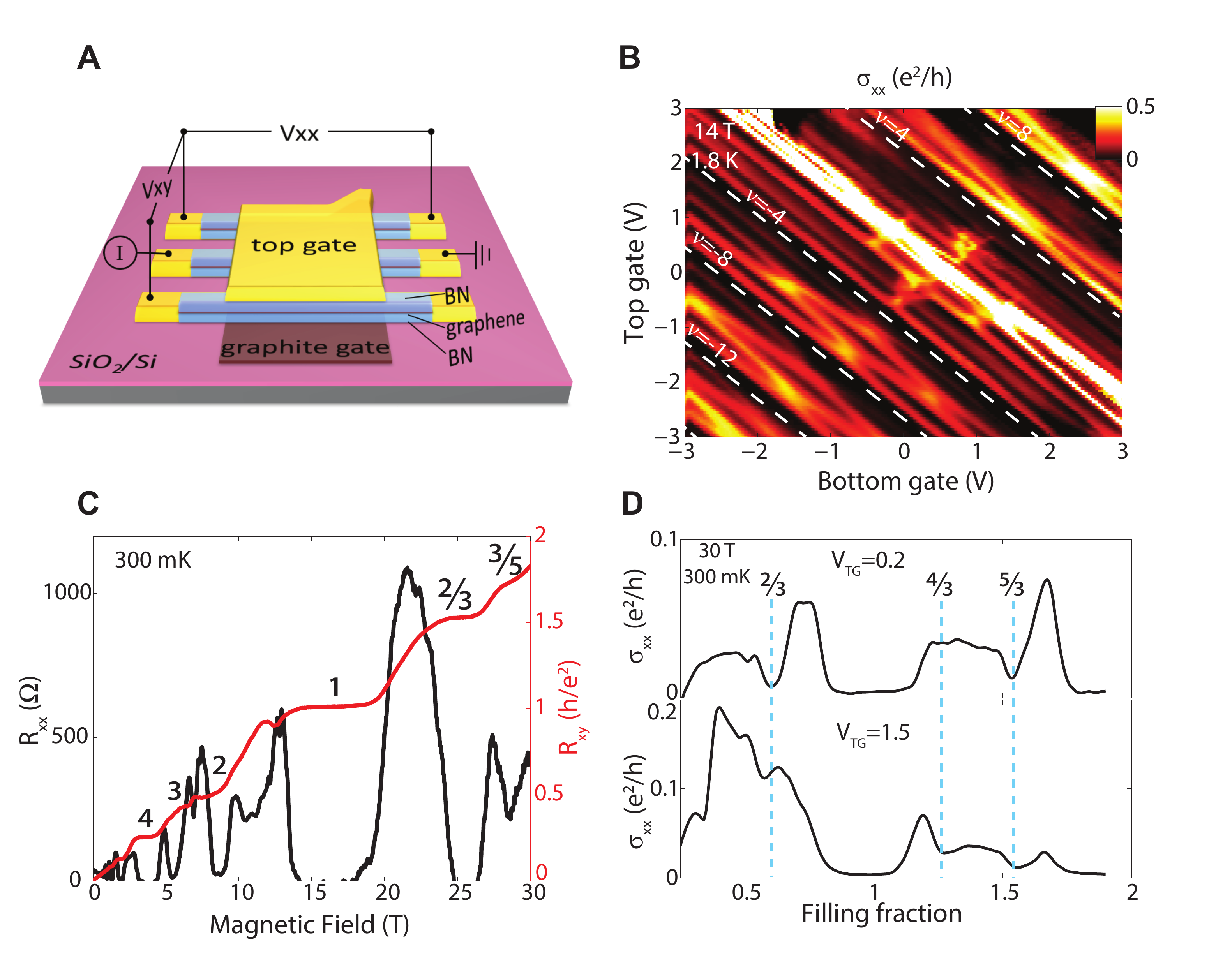}
	\end{center}
\end{figure}

\noindent {\bf Fig. 1.} {\bf A)} Schematic diagram of the dual-gate device architecture. {\bf B)} $\sigma_{xx}$ as a function of top and bottom gate voltages at B=14~T, and T=1.8~K. All broken symmetry integer states are visible. White dashed lines indicate degenerate cyclotron gaps {\bf C)} $R_{xx}$ and $R_{xy}$ as a function of magnetic field at a fixed carrier density ($n=4.2\times10^{11}$~cm$^{-2}$). A fully developed $\nu=2/3$ state appears at $\sim$25 T, with a 3/5 state developing at higher field. {\bf D)} $\sigma_{xx}$ vs. filling fraction at 30~T, 300~mK acquired by sweeping the bottom gate for two different top gate voltages.

\clearpage

\begin{figure}[t]
    \begin{center}
	\includegraphics[width=\linewidth]{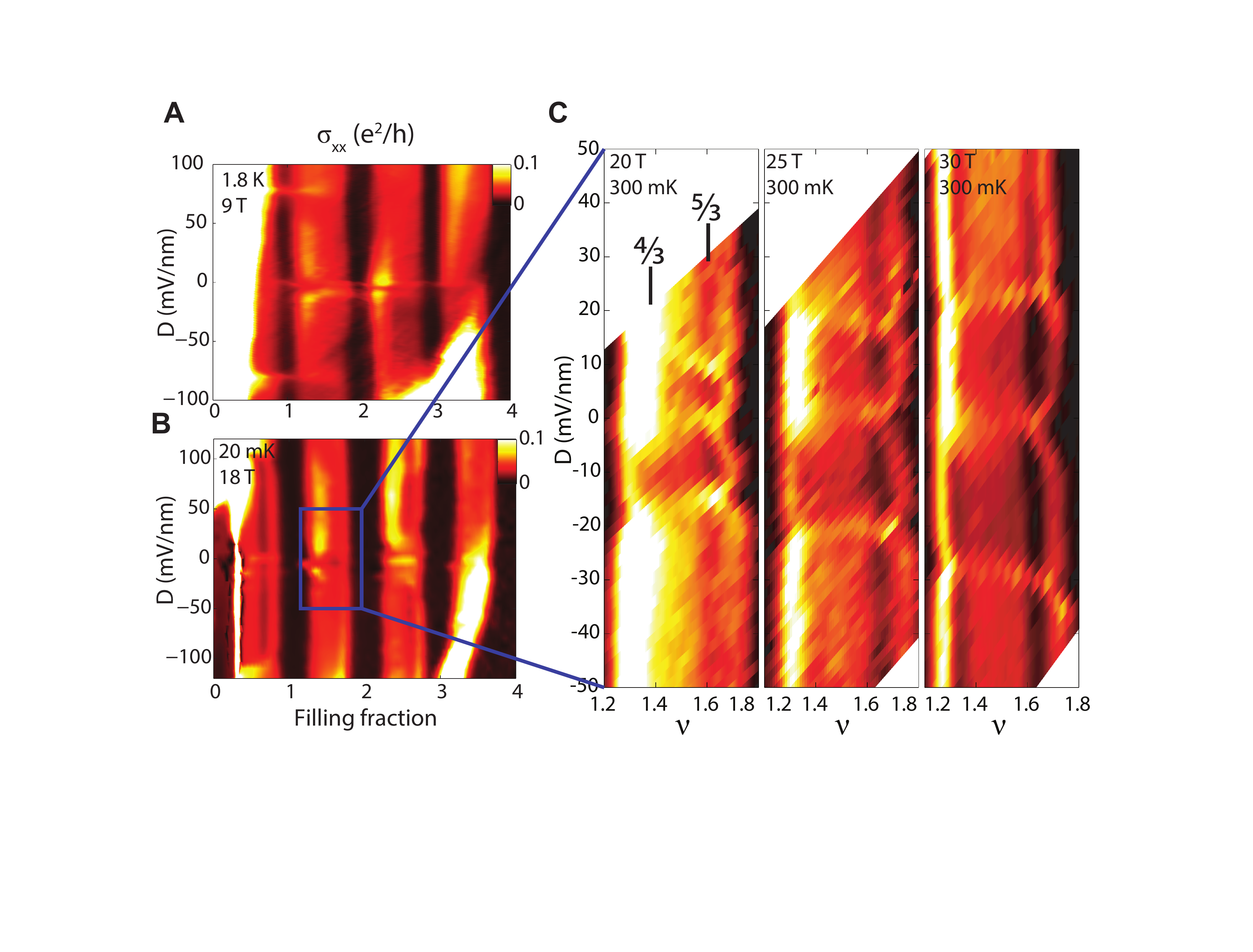}
	\end{center}
\end{figure}

\noindent {\bf Fig. 2.} {\bf A)} $\sigma_{xx}$ at $B=9$~T and $T=1.5$~K vs. filling fraction and displacement field showing full integer symmetry breaking for 1$\leq \nu \leq$4. {\bf B)} $\sigma_{xx}$ at $B=18$~T and $T=20$~mK showing both  well developed minima at fractional filling factors, and clear transitions with varying displacement field.  {\bf C)} High resolution scans of the region 1$<\nu<$2, -50$<D<$50~mV/nm at 300~mK for three different magnetic fields . 

\clearpage

\begin{figure}[t]
    \begin{center}
	\includegraphics[width=\linewidth]{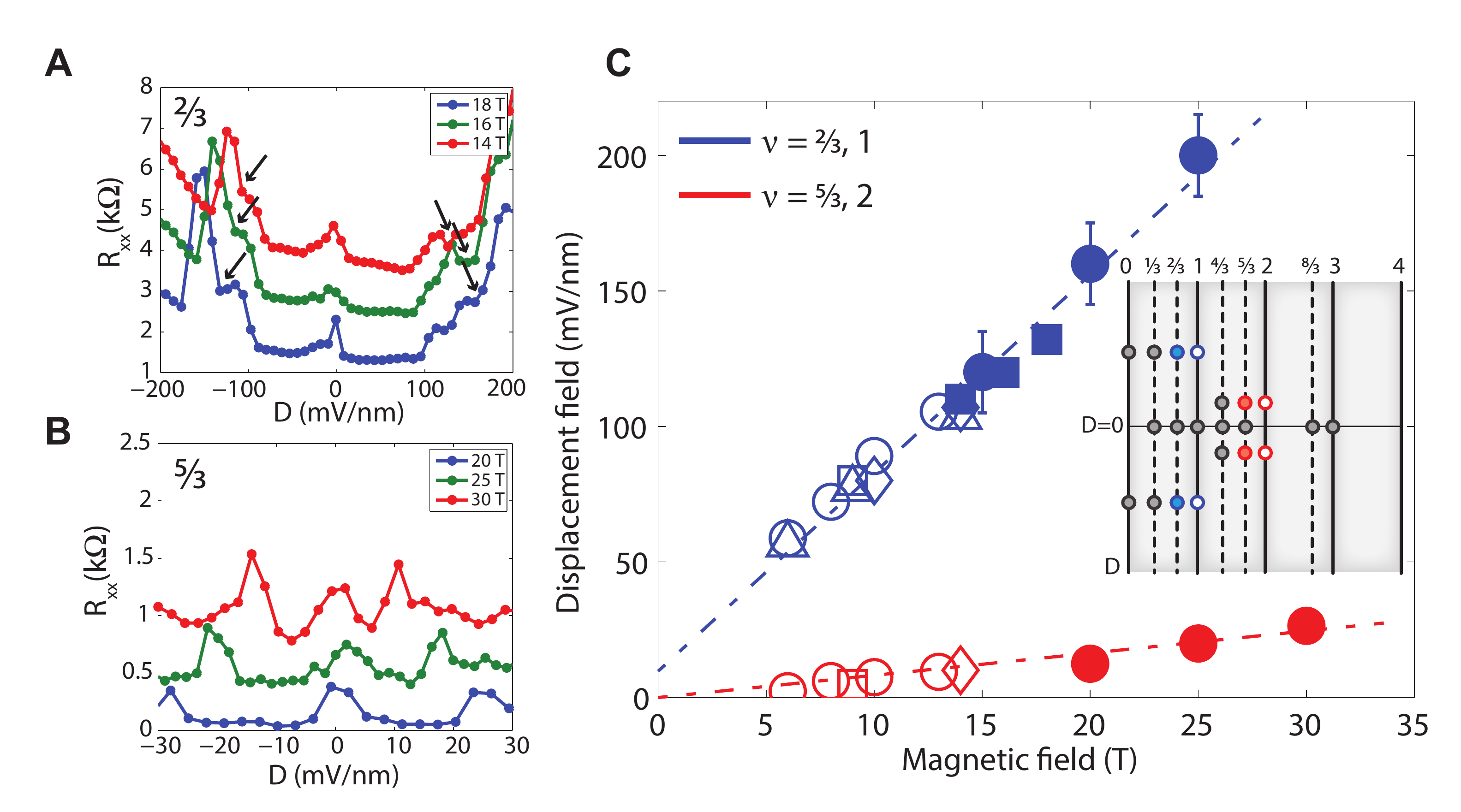}
	\end{center}
\end{figure}

\noindent {\bf Fig. 3.} {\bf A)} $R_{xx}$ versus displacement field at filling fraction $\nu=2/3$ acquired at three different magnetic fields. Arrows mark middle of transition. Data are offset for clarity {\bf B)} Similar data as in a but corresponding to $\nu=5/3$  {\bf C)} Plot of all observed non-$D=0$ transitions for $\nu=2/3$, 1, 5/3, and 2. $\nu=2/3$ and 1 are blue, $\nu=5/3$ and 2 are red. Open symbols correspond to integers, filled symbols to fractions. The data was acquired from 4 different devices, with each shape corresponding to a unique device. Error bars are not shown where they are smaller than the markers. Inset is a schematic map of the transitions in the symmetry broken QH states for $0\leq\nu\leq4$. Vertical lines correspond to the QH state for the filling fraction at the top. Boxes located on the central horizontal line indicate transitions at $D=0$ and boxes located away from the horizontal line indicate transitions at finite $D$. Filled and empty colored circles represent the transitions in the main panel while the black circles represent other transitions presented in Fig. 2.



\section*{Supplementary material}

# Supplementary Materials for
**Tunable Fractional Quantum Hall Phases in Bilayer Graphene**

Patrick Maher[1†], Lei Wang[2†], Yuanda Gao[3], Carlos Forsythe[1], Takashi Taniguchi[4], Kenji Watanabe[4], Dmitry Abanin[5,6], Zlatko Papić[5,6], Paul Cadden-Zimansky[7], James Hone[3], Philip Kim[1*], and Cory R. Dean[8*]

[1]Department of Physics, Columbia University, New York, USA
[2]Department of Electrical Engineering, Columbia University, New York, NY, USA
[3]Department of Mechanical Engineering, Columbia University, New York, NY, USA
[4]National Institute for Materials Science, 1-1 Namiki, Tsukuba, Japan
[5]Perimeter Institute for Theoretical Physics, Waterloo, ON N2L 2Y5, Canada
[6]Institute for Quantum Computing, Waterloo, ON N2L 3G1, Canada
[7]Physics Program, Bard College, Annandale-on-Hudson, NY 12504, USA
[8]Department of Physics, The City College of New York, New York, NY, USA

[†]These authors contributed equally to this work
[*]Corresponding authors. E-mail: cdean@ccny.cuny.edu, pk2015@columbia.edu

**This PDF file includes:**

    Materials and Methods
    Supplementary Text
    Figs. S1 to S9
    References

Contents

1. Materials and Methods

2. Supplementary Text

References

# 1. Material and Methods

**1.1 Device fabrication process**

The dual-gated bilayer graphene device is fabricated in the process shown in Fig. S1. First, graphene, BN, and graphite flakes are each prepared on a piranha-cleaned surface of $SiO_2$/Si by mechanical exfoliation. An optical microscope is used to identify flakes with the proper thickness and layer number. The surface of every flake is then examined by an atomic force microscope (AFM) in non-contact mode. Only flakes which are completely free from particles, tape residues and step edges are selected. The stack is assembled using the same method as in our previous work, the van der Waals (vdW) transfer method (*28*). As shown in Fig. S1A, B, C, the BN/graphene/BN stack is made using the top BN flake to pick up a bilayer graphene flake and the bottom BN flake in sequence. The stack is then transferred on a flake of graphite (Fig. S1E), which is used as the local gate. Fig. S2A shows an AFM image of the BN/graphene/BN/graphite stack, which is totally bubble free. The stack is then etched to expose the graphene edges, which are outside of the graphite region (Fig. S1F). Edge-contact is then made by depositing 1/10/60 nm of Cr/Pd/Au layers. A metal top gate is finally deposited to achieve the dual gated device (Fig. S2B).

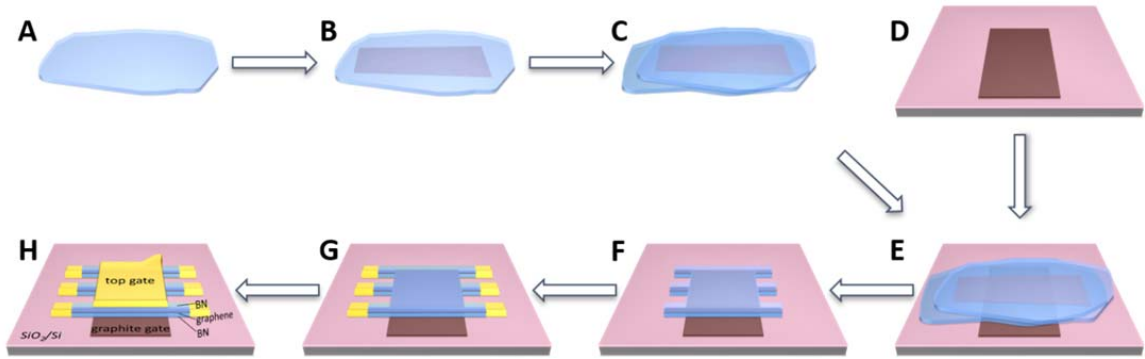

**Fig. S1. Schematic of the fabrication process of a dual-gated BN/graphene/BN device.** (A) Top layer h-BN flake (PPC film is not shown in this schematic). (B) Top BN flake picks up a piece of bilayer graphene using the vdW method. (C) Top BN flake with bilayer graphene picks up bottom layer BN using the vdW method. (D) A piece of graphite, which will be used as the local gate, is prepared on $SiO_2$/Si substrate by mechanically exfoliation. (E) The BN/graphene/BN stack is transferred onto the graphite using the vdW method. (F) The BN/graphene/BN stack is etched. (G) Electrical leads to the device are made by edge contact. (H) Top gate metal is deposited.

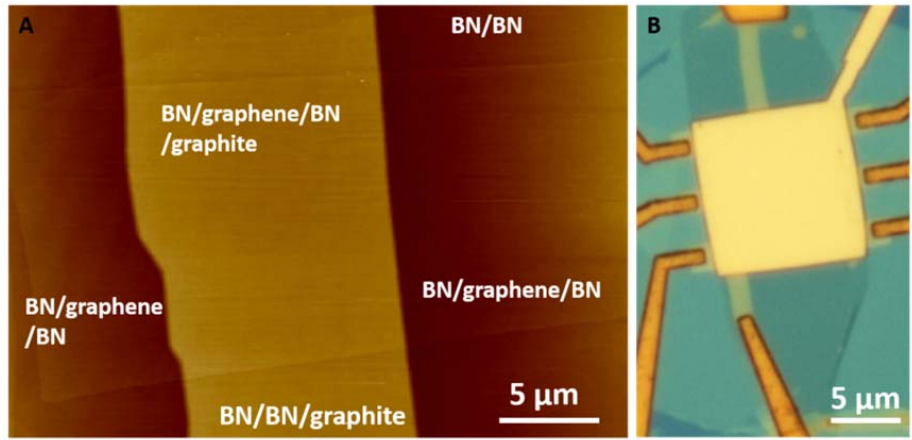

**Fig.S2. AFM and optical images of the device.** (A) AFM image of the BN/graphene/BN/graphite stack on $SiO_2$ substrate, taken at the stage of Fig.S1E step. (B) Optical image of the final device.

## 2. Supplementary text

### 2.1 Low temperature transport at zero magnetic field

Fig. S3 shows the measured resistivity as a function of the local graphite gate voltage. The bottom BN thickness is 18.5 nm measured by AFM. Negative resistance is observed, indicating ballistic transport across the diagonal of the device. That gives an estimation of the lower bound of the carrier mean free path to be 15 μm, which corresponds to a carrier mobility value of 1 000 000 cm$^2$/Vs.

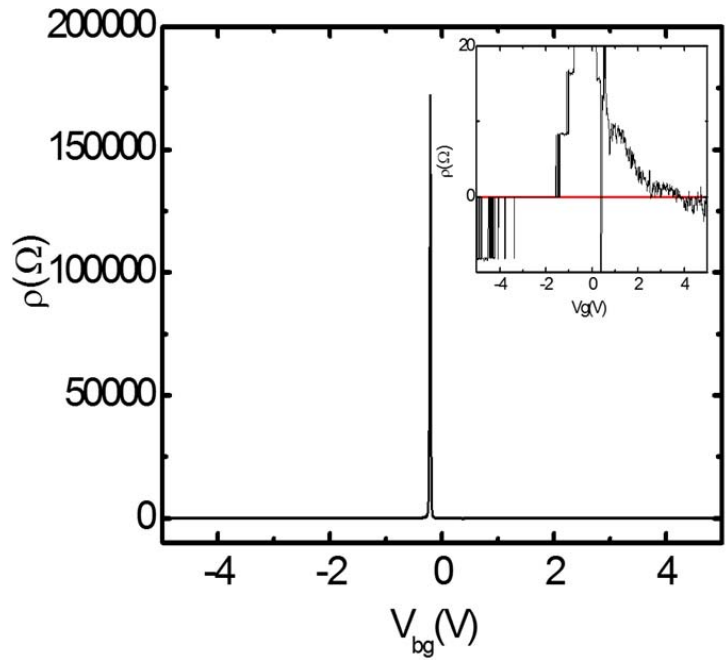

**Fig. S3. Four terminal resistivity measured as a function on local graphite gate voltage.** Zooming-in on the curve near zero resistance is shown in the inset. A negative resistance is observed, indicating ballistic transport.

**2.2 Effect of gate tuning the graphene leads on the quantum Hall measurement**

In all our devices, we keep a distance (typically ranging from 0.5 to 2 μm, depending on the alignment accuracy of the lithography) between the metal leads and the top and bottom gate to prevent any direct short or leakage between the device layer and the gates. This results in a small length of the lead which is made of graphene, and which is not dual gated by the local top and bottom gates (Fig. S4). Due to graphene's low resistivity, these graphene leads are completely suitable voltage probes at zero magnetic field. At high magnetic field, however, the $v = 0$ state is strongly insulating (*37*). Thus any PN junction taking place in the graphene lead will become highly resistive and will result in a poor voltage probe. To overcome this problem, a high voltage is applied to the silicon back gate to induce a high carrier density in the graphene leads of the same type as the device channel. The device channel is not affected by the silicon back gate thanks to the screening of the local graphite gate. Fig. S5 shows a comparison of quantum Hall measurements of the longitudinal resistance $R_{xx}$ and the Hall conductivity $\sigma_{xy}$ as a function of local back gate voltage, with the silicon gate being set to 0 V and 60 V. The local top gate is set to 0 V. Clearly, on the electron side, better developed plateaus in $\sigma_{xy}$ and minima in $R_{xx}$ are observed when the silicon gate is set to 60 V. However, on the hole side data are not improved due to a PN junction between the lead and the channel. Conversely, when the silicon gate is set to be – 60 V, the hole side is improved and the electron side is degraded.

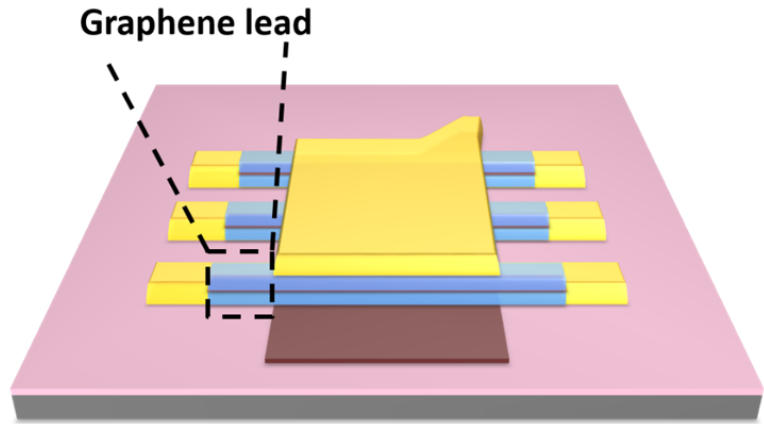

**Fig. S4. Schematic diagram showing the graphene leads of the device.**

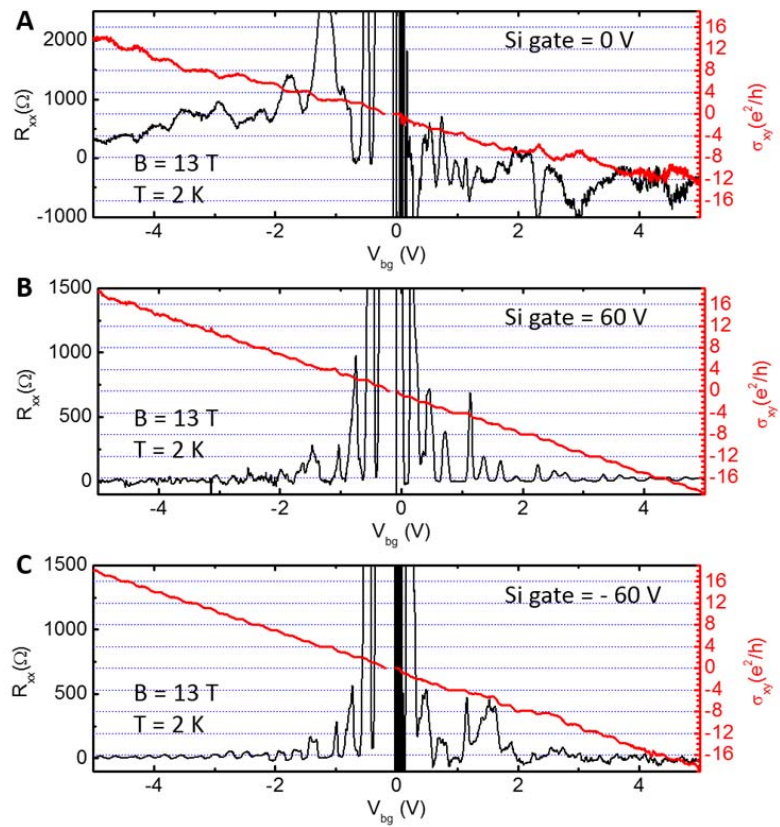

**Fig. S5. Comparison of the quantum Hall measurements for different induced carrier densities in the graphene leads:** (A) silicon gate voltage is 0 V, (B) silicon gate voltage is 60 V, (C) silicon gate voltage is – 60V.

7

## 2.3 Longitudinal and Hall conductivity versus carrier density at 31T, 300mK

The longitudinal and Hall conductivity are measured under a magnetic field of 31 T at 300 mK. In Fig. S6, a detailed plot of the lowest Landau level is given. We observe the quantization of $\sigma_{xy}$ to values of $\nu e^2/h$ concomitant with minima in $R_{xx}$ at fractional filling factors $\nu$ = -10/3, -7/3, -4/3, -1/3, 1/3, 2/3, 5/3, and 8/3. This sequence breaks electron-hole symmetry (*14*) and seems to indicate that the sequence described by $\nu$ = m-1/3, where m is an integer, are the most robust fractional states.

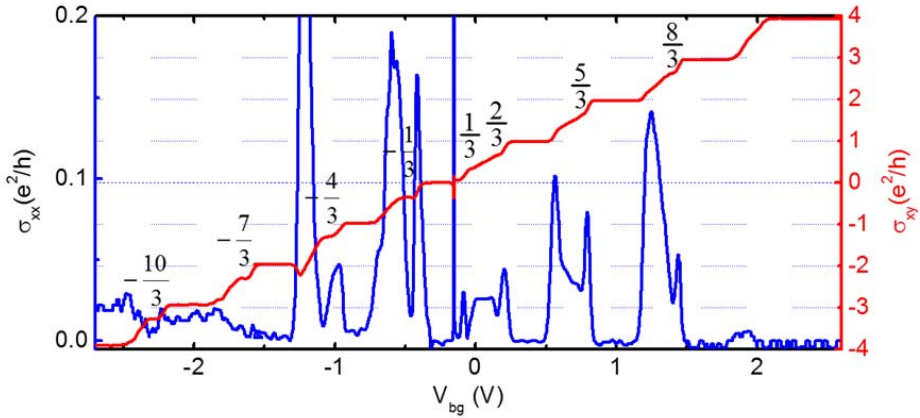

**Fig. S6. Longitudinal and Hall conductivity as a function of local back gate.**

## 2.4 Fractional quantum Hall state ν=2/3 as a function of displacement field

As shown in Fig. S7, the finite D transition for $\nu$ = 2/3 state occurs at higher D values compared with $\nu$ = 5/3 state. As the magnetic field increases, the transition broadens and its D value also increases.

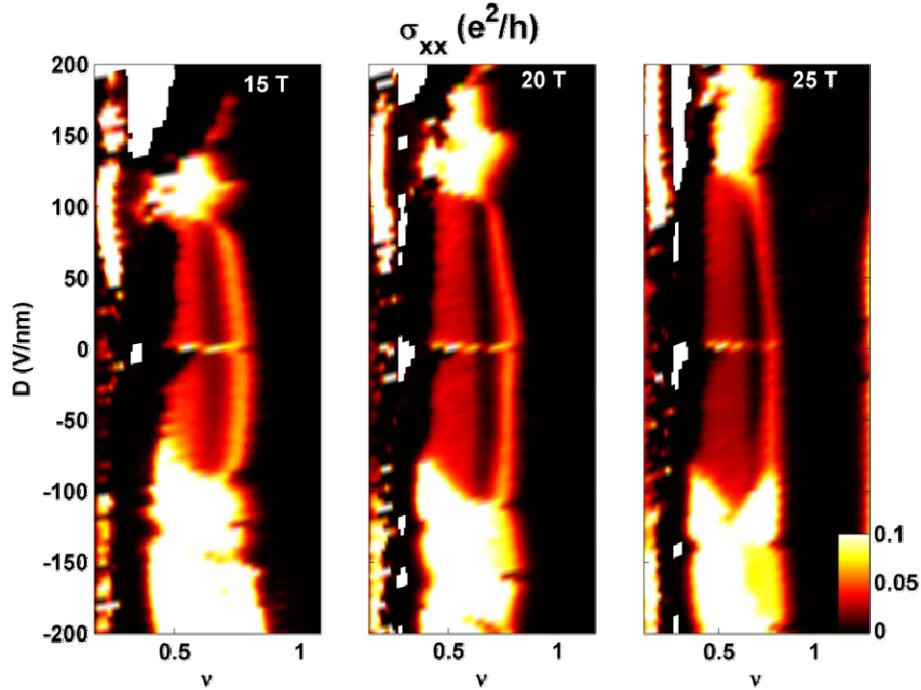

**Fig. S7. Fractional quantum Hall state 2/3 as a function of displacement filed under magnetic field values of 15 T, 20 T and 25 T**

## 2.5 Fractional quantum Hall for ν>4

In addition to the fractional quantum Hall states in the range -4<ν<4, we also observe fractional states in the range 4<ν<8. We see evidence of 14/3, 17/3, 20/3, and 23/3, all of which follow the n-1/3 hierarchy proposed in section 2.3 (Fig. S7 a). We do not see obvious transitions in displacement field in these higher LL fractions, but this could well be due to measurement issues related to the extremely high gate voltages needed to observe these states (Fig. S7 b). We observe no evidence of fractions for ν>8.

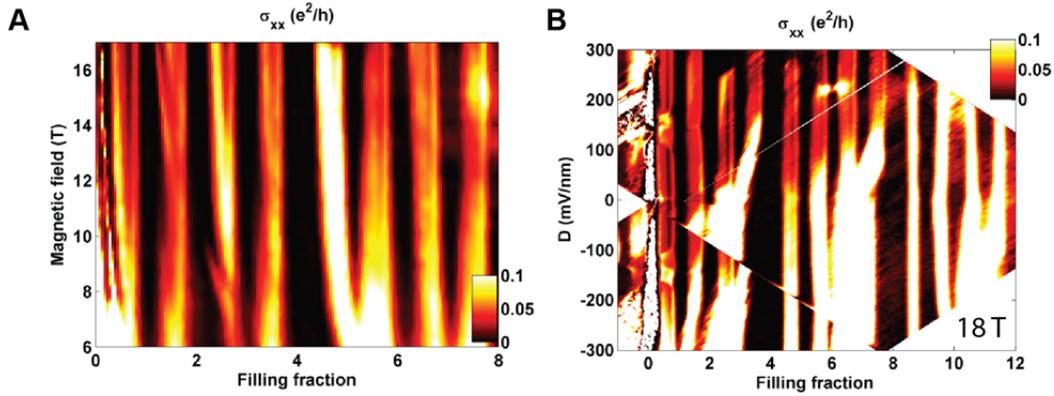

**Fig. S8. Fractional quantum Hall in higher Landau levels:** A) Longitudinal conductivity as a function of filling fraction and magnetic field. Data taken at 20 mK. B) Longitudinal conductivity as a function of filling fraction and displacement field at 18 T, 20 mK.

## 2.6 Data sets for Fig. 3c

In figure S9 we display the remaining data that went in to Fig. 3c. Data from the device marked as a circle is also presented in Fig. 2c, Fig. 3b, and Fig. S7. Data from the device marked as a square is also in Fig. 2a, Fig. 2b, Fig. 3a, and Fig. S8. Data from the device marked as a diamond is also presented in Fig. 1b.

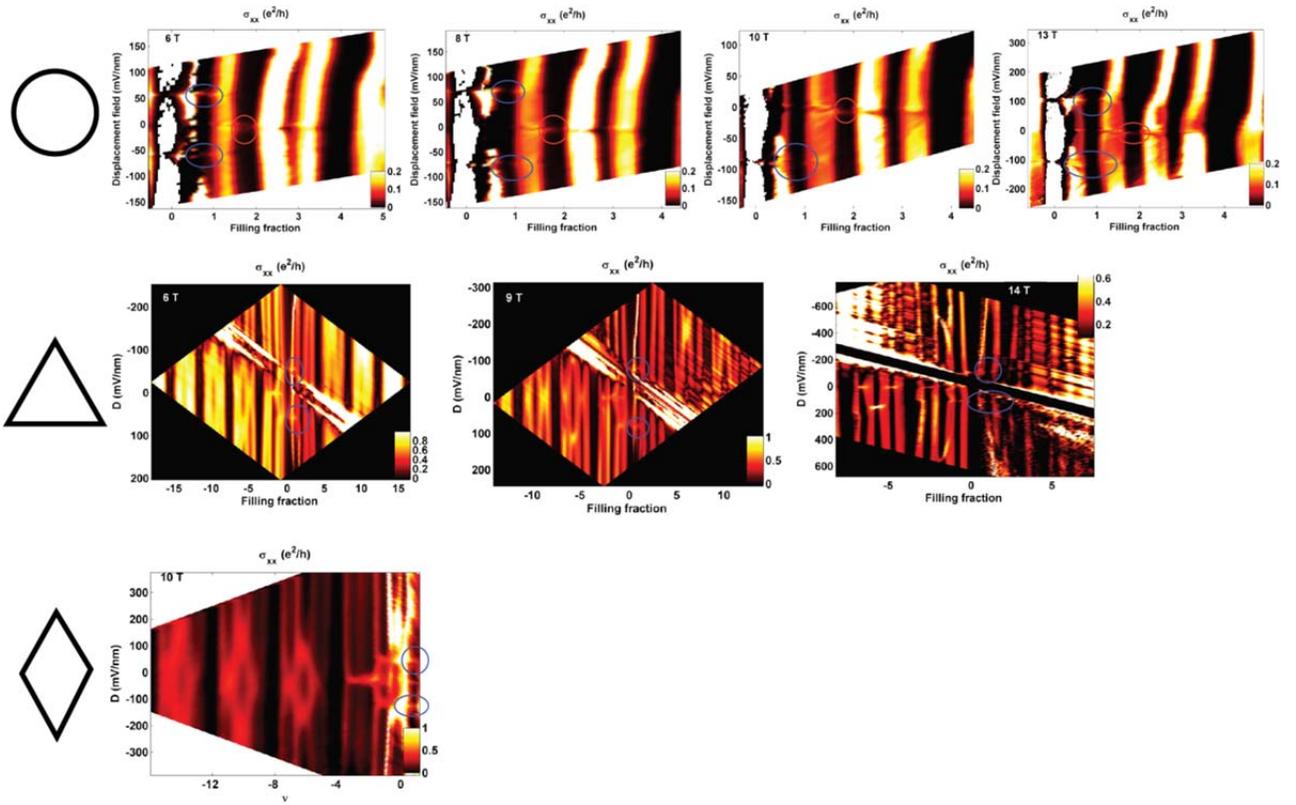

**Fig. S9. All previously unpresented datasets for the plot in Fig. 3c:** Symbols designate different devices and correspond to symbols in Fig. 3c. Transitions for ν=1 (ν=2) are circled in blue (red).

**References**

37. Feldman, B. E., Martin, J. & Yacoby, A. *Nat Phys* **5,** 889–893 (2009).



\end{document}